\documentclass[twocolumn,english,aps,graphicx,amsmath,showpacs]{revtex4}
\usepackage[T1]{fontenc}
\usepackage[latin1]{inputenc}
\usepackage{babel}
\usepackage{graphics}

\makeatletter

\providecommand{\LyX}{L\kern-.1667em\lower.25em\hbox{Y}\kern-.125emX\@}

\makeatother
\begin{document}

\title{Stable isochronal synchronization of mutually coupled chaotic lasers}

\author{Einat Klein\( ^{1} \), Noam Gross\( ^{1} \), Michael Rosenbluh\( ^{1} \),
Wolfgang Kinzel\( ^{2} \), Lev Khaykovich\( ^{1} \), Ido
Kanter\(^{1} \)}

\affiliation{\( ^{1} \)Department of Physics, Bar-Ilan University,
Ramat-Gan, 52900 Israel,}

\affiliation{\( ^{2} \)Institut f\"ur Theoretische Physik,
Universit\"at W\"urzbur, Am Hubland 97074 W\"urzburg, Germany}

\begin{abstract}
The dynamics of two mutually coupled chaotic diode lasers are
investigated experimentally and numerically. By adding self
feedback to each laser, stable isochronal synchronization is
established. This stability, which can be achieved for symmetric
operation, is essential for constructing an optical public-channel
cryptographic system. The experimental results on diode lasers are
well described by rate equations of coupled single mode lasers.
\end{abstract}




\maketitle A semiconductor laser, when subjected to optical
feedback, displays chaotic behavior
\cite{RouteFeedback,CoexistenceFeedback,LowFeedback}. This
phenomenon has been investigated over the last two decades with
synchronization between two chaotic lasers attracting much recent
interest  because of its applicability to secret communication
\cite{SynchronizationEncryption,SynchronizedRates,Donati,Larger}.
Different configurations, such as delayed optoelectronic feedback
\cite{ChaosFeedback,ExperimentalLasers,NonlinearCoupling} or
coherent optical injection
\cite{ChaosLasers,ExperimentalFeedback,SynchronizationLasers} have
been suggested for synchronization of two semiconductor lasers.
The coupling between the lasers is accomplished in a
unidirectional
\cite{ChaosFeedback,SynchronizedRates,ExperimentalLasers} or
bidirectional
\cite{ChaosLasers,NonlinearCoupling,ExperimentalFeedback} fashion,
leading to different kinds of synchronization phenomena. Recently
it was shown that it might be possible to use the synchronization
of two symmetric chaotic lasers for novel cryptographic
key-exchange protocols, whereby secret messages can be transmitted
over public channels without using any previous secrets
\cite{Kanter}. A necessary condition for such a protocol is the
symmetry in the configuration: Two identical chaotic lasers should
be coupled by identical mutual interactions.

In a face to face laser configuration (mutual coupling) where the
setup is built symmetrically, isochronal synchronization, however,
was always found to be unstable and one laser had to be slightly
detuned to guarantee a well defined leader/laggard configuration
in order to achieve high fidelity synchronization
\cite{ChaosLasers,SynchronizationLasers,White}.

\begin{figure}

{\centering \resizebox*{0.5\textwidth}{0.2\textheight}
{{\includegraphics{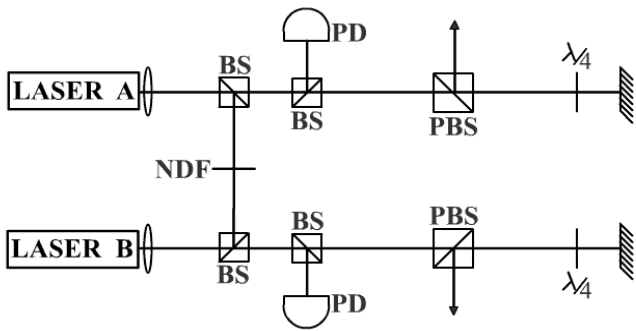}}}
\par}
\caption{\label{Schema}A schematic figure of the two coupled
lasers. BS - Beam Splitter; PBS - Polarization Beam Splitter; NDF
- Neutral Density Filter; PD - Photodetector}
\end{figure}

In this paper, we present a novel configuration of two symmetric
lasers that exhibit stable isochronal synchronization under
symmetric operation conditions. The two lasers take equal roles in
creating and maintaining synchronization without any symmetry
breaking. Although message transfer using this system has yet to
be tested, this result is an important ingredient in the
transmission of secret messages over a single public channel in
both directions.

A schematic of our experimental system is shown in Fig.
\ref{Schema}. Each laser receives a delayed signal from the other
as well as a delayed self-feedback. The time delay between the
lasers is denoted $\tau_{c}$, the delay of the self feedback is
denoted $\tau_{d}$, and the  coupling and self feedback rates are
denoted $\sigma$ and $\kappa$ respectively. The results presented
in this paper are for $\tau_{d}=\tau_{c}=7ns$ but synchronization
was observed for other time delays as well. We measure the degree
of synchronization by the time-dependent cross correlation
\cite{UnidirectionallyLasers}, which is denoted as $\rho$ and
defined as follows:

\begin{equation}\label{eqrho}
    \rho(\Delta t) = \frac{\sum_i(I_{A}^{i}-<I_{A}^{i}>)\cdot{(I_{B}^{i+\Delta t}-<I_{B}^{i+\Delta t}>})}{\sqrt[]{\sum_i(I_{A}^{i}-<I_{A}^{i}>)^{2}\cdot\sum_i(I_{B}^{i+\Delta t}-<I_{B}^{i+\Delta t}>)^{2}}}
\end{equation}

\noindent $I_{A}$ and $I_{B}$ are the time dependent intensities
of lasers A and B respectively. We found it convenient to perform
the experiments in a synchronization regime where total laser
intensity breakdowns take place, commonly referred to as the LFF
(Low Frequency Fluctuation)
\cite{RouteFeedback,CoexistenceFeedback,LowFeedback} regime. In
this regime, synchronization as evidenced by the correlated
intensity breakdown of both lasers is easily observed.  During the
intensity breakdown, however, the system can temporarily
desynchronize, as was proven numerically in ref.
\cite{HyperchaoticLasers}. In order to avoid these irregularities
in our data analysis we divide the sequences into segments and  we
exclude segments containing an LFF breakdown. The correlation
coefficient is calculated between matching time segments and then
averaged over all segments. The time scale of the intensity
fluctuations was of the order of 1ns, which is also the
experimental time resolution. For this reason we also averaged the
simulation results over 1ns , so that $I_{A}^{i}$ is an averaged
intensity for a window of 1ns at time $i$. We arbitrarily chose
the size of each segment to be 10ns, which is an order of
magnitude smaller than the average time between breakdowns -
around 150ns, and thus each segment consists of ten points (each
point of 1ns).

\begin{figure}

{\centering \resizebox*{0.5\textwidth}{0.38\textheight}
{{\includegraphics{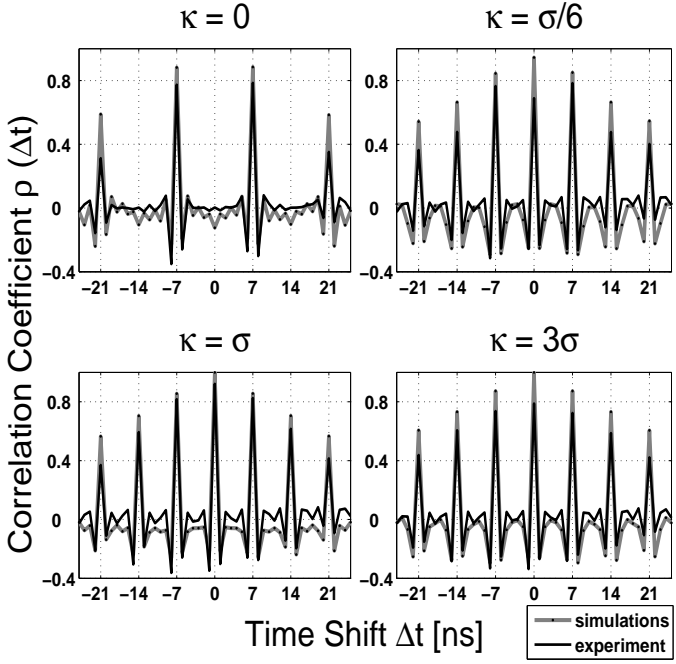}}}
\par}
\caption{\label{fb_hist} Numerical and experimental results.
Correlation coefficient between laser intensities at different
time delays for $\tau_{c}=\tau_{d}=7ns$. $\kappa$ - self feedback
rate. $\sigma$ - coupling rate. Different $\kappa$ values affect
the dynamics of the synchronization (in the simulations we kept
$\kappa + \sigma=10^{11}$). For $\kappa$ = 0 (face to face
configuration) we observe a coexistence of leader/laggard
situation. When $\kappa>0$ the isochronal synchronization is
stable (maximum correlation in zero time delay).}
\end{figure}

Fig. \ref{fb_hist} shows the shifted correlation coefficient
$\rho(\Delta t)$ \cite{ExperimentalLasers}, which is obtained by
calculating the correlation coefficient between the outputs of the
two lasers when one is continuously shifted in time with respect
to the other. The four graphs exhibit the dynamics of
synchronization for different relations between coupling and self
feedback rates, $\sigma$ and $\kappa$ \cite{remark2}.


In all the graphs the symmetry is evident. Without self-feedback,
when $\kappa=0$, we find high correlation for time delays of
$\pm\tau_{c}$ (7ns) and no correlation at zero time-delay,
implying an achronal synchronization that was discussed in Ref.
\cite{SynchronizationLasers}. Further investigation of the
symmetry of the achronal synchronization is given later in this
paper.

One can clearly observe that in the other three cases displayed,
for which $\kappa>0$, there appears a very high correlation at
zero time-delay. For $\kappa=\sigma$ for instance, the experiment
results show an average value of 0.92 and a most probable value of
0.99. In the simulations we report an even higher correlation
which indicates complete synchronization in between the
breakdowns. We can also see secondary peaks at $\Delta t = \pm n
\cdot\tau _{d}$ where $n$ is an integer. These peaks reflect the
fact that the chaotic waveform has some self correlation at time
intervals of $n\cdot\tau _{d}$ where $n$ is a small integer, due
to the self-feedback. The above results hold, experimentally and
in simulations, for the case of $\tau _{c} = \tau _{d}$. An
interesting observation is that also for $\tau _{c} = n\cdot\tau
_{d}$ where $n$ is a small integer (up to about 3), stable
isochronal synchronization appears. This is supposedly because of
the self-correlation mentioned above.

\begin{figure}

\vspace{-0.8 cm} {\centering
\resizebox*{0.5\textwidth}{0.3\textheight}
{{\includegraphics{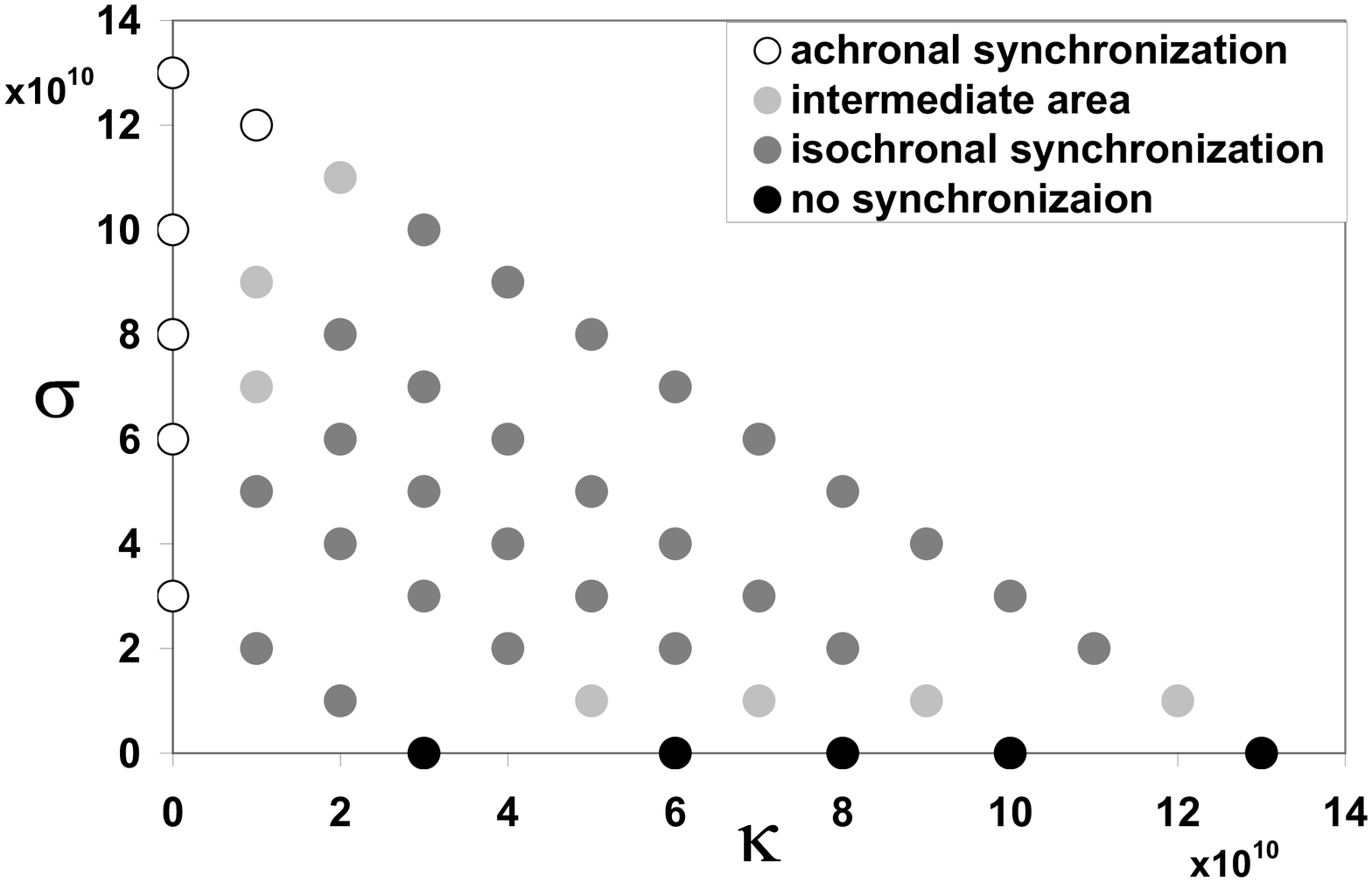}}}
\par}
\vspace{-0.6 cm} \caption{\label{PhaseSpace}The phase space of
$\kappa$ and $\sigma$, with $\kappa_{A}=\kappa_{B}$,
$\sigma_{A}=\sigma_{B}$ and $\tau_{c}=\tau_{d}$ \cite{remark1}.}
\end{figure}

In our numerical simulations we explored the phase space of
$\kappa$ and $\sigma$ for $\tau _{c} = \tau _{d}$ as displayed in
Fig. \ref{PhaseSpace}. Stable isochronal synchronization appears
over a wide range of values of $\kappa$ and $\sigma$ (the dark
grey circles in the graph). Without self-feedback the isochronal
solution is unstable and achronal synchronization appears
indicated by the open circles. Achronal synchronization also
appears when $\tau_{c}\ne\tau_{d}$.


We now give a more detailed description of the numerical
simulations and the experimental results. To numerically simulate
the system we used the Lang-Kobayashi equations \cite{Kobayashi}
that are known to describe a chaotic diode laser. The dynamics of
laser $A$ are given by coupled differential equations for the
optical field, $E$, the time dependent optical phase, $\Phi$, and
the excited state population, $n$;

\begin{eqnarray}
\frac{dE_{A}}{dt}=\frac{1}{2}G_{N}n_{A}E_{A}(t)+\frac{C_{sp}\gamma[N_{sol}+n
_{A}(t)]^{-1}}{2E_{A}(t)} ~~~~~~~~~~~~~~~~~~~~~~~~~~~ &  &
\nonumber\\
+ \kappa_{A} E_{A}(t-\tau _{d})cos[\omega _{0} \tau + \Phi
_{A}(t)-\Phi_{A} (t-\tau
_{d})]~~~~~~~~~~~~~~~~~\nonumber\\
+ \sigma_{A} E_{B}(t-\tau _{c})cos[\omega _{0} \tau _{c}
+\Phi_{A}(t)-\Phi
_{B}(t-\tau _{c})]~~~~~~~~~~~~~~\nonumber\\
\frac{d\Phi_{A}}{dt}=\frac{1}{2}\alpha G_{N}n_{A}
~~~~~~~~~~~~~~~~~~~~~~~~~~~~~~~~~~~~~~~~~~~~~~~~~~~~~~~~~~~~~~ & &
\nonumber\\
-\kappa_{A} \frac{E_{A}(t-\tau_{c})}{E_{A}(t)}sin[\omega _{0} \tau
+\Phi _{A}(t)-\Phi_{A}
(t-\tau _{d})] ~~~~~~~~~~~~~~~~~ &  & \nonumber\\
-\sigma_{A} \frac{E_{B}(t-\tau _{c})}{E_{A}(t-\tau
_{c})}sin[\omega _{0} \tau _{c} +\Phi_{A}(t)-\Phi _{B}(t-\tau
_{c})]~~~~~~~~~~~~~~~\nonumber\\
\frac{dn_{A}}{dt}=(p-1)J_{th}-\gamma n_{A}(t)-[\Gamma +
G_{N}n_{A}(t)]E_{A}^{2}(t)~~~~~~~~~~~~~~~~~~~ &  &
\nonumber\\
\nonumber\end{eqnarray}

\noindent and likewise for laser B. The values and meaning of the
parameters are those used in Ref. \cite{HyperchaoticLasers}.

In the experimental setup we use 2 device-identical single-mode
semiconductor lasers emitting at 660 nm and operated close to
their threshold currents. The temperature of each laser is
stabilized to better than 0.01K. Both lasers are subjected to
similar optical feedback. The length of the external cavities is
equal for both lasers and is set to 105 cm (round trip time
$\tau_{d}$ = 7 ns). The feedback strength of each laser is
adjusted using a $\lambda/$4 wave plate and a polarizing beam
splitter and is set to approximately 2$\%$ of the laser's power.
It leads to a reduction of about 5$\%$ in the solitary laser's
threshold current. The two lasers are mutually coupled by
injecting another 2$\%$ of each one's output power into the other
one. Coupling power is controlled by a neutral density filter. The
coupling optical path is set to 210 cm ($\tau_{c}$ = 7 ns) which
is equal the round trip length in the external cavities
($\tau_{c}$ = $\tau_{d}$) . Two fast photodetectors (with response
time $<$ 500 ps) are used to monitor laser intensities which are
simultaneously recorded with a digital oscilloscope (500 MHz ,1
GS/sec). Fine frequency tuning of the two lasers is crucial for
the establishment of synchronization. This can be achieved by
scanning one laser's temperature. While doing so we monitor both
laser intensity signals on the digital scope. The desired
temperature is attained when there is no obvious leader or
laggard, i.e. both breakdowns occur simultaneously or within a
time less than $\tau_{c}$ of each other and neither laser can be
declared as the leader. A 0.02K deviation from this temperature
the symmetry breaks and one of the lasers turns into a laggard or
leader, i.e. it's signal seems to follow or precede the other
laser's signal by a time $\tau_{c}$. Whether the laser becomes a
leader or a laggard depends very sensitively on the relative
output powers of the two lasers.

\begin{figure}

{\centering \resizebox*{0.5\textwidth}{0.29\textheight}
{{\includegraphics{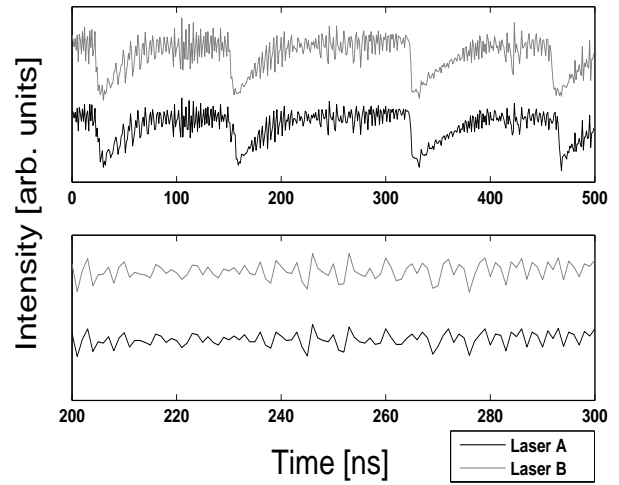}}}
\par}
\caption{\label{trace} A typical experimental time sequence of
laser intensities for isochronal synchronization. The system
parameters are $\kappa$ = $\sigma$ and $\tau_{c}$ = $\tau_{d}$.
The lower figure is a more detailed view of the figure above it.}

\end{figure}

Isochronal synchronization is established when the self feedback
round trip times ($\tau_{d}$) of both lasers and the coupling
optical travel time ($\tau_{c}$) between them are all equal. In
this type of synchronization we get maximum overlap between the
two signals for zero time delay between them. For fine tuning of
the optical path lengths, two of the mirrors in the experimental
setup were mounted on translation stages to simultaneously adjust
$\tau_{dA}$ and $\tau_{c}$ to be equal to each other and to
$\tau_{dB}$ which remained constant. We found that once high
fidelity synchronization was established, even a change of 100
$\mu$m in the location of the mirrors caused the synchronization
to deteriorate. This agrees with the results of the simulation
that show that the synchronization is sensitive to small changes
in $\tau_{c}$ and $\tau_{d}$.

\begin{figure}

{\centering \resizebox*{0.5\textwidth}{0.3\textheight}
{{\includegraphics{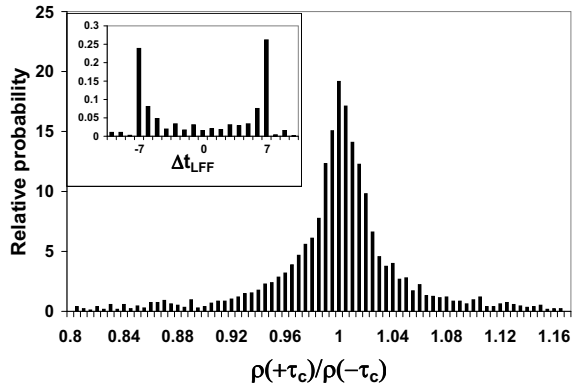}}}
\par}
\vspace{-0.5 cm} \caption{\label{histo_achronal} Numerical
simulations results for the achronal synchronization with
$\kappa=0$, $\sigma=10^{11}$ and $\tau_{c}=7ns$. The relative
probability of the ratio of the cross correlation of two time
delays $\rho(+\tau_{c})$ and $\rho(-\tau_{c})$. Inset: A histogram
of the time delay between the breakdowns of the two lasers $A$ and
$B$, for the same parameters as above.}

\end{figure}

One of the challenging applications of chaotic lasers is in
cryptographic systems. Unidirectional coupling was used for the
creation of a secret-key cryptographic method \cite{UniCrypto},
however, for the purpose of public-key systems such as a
key-exchange protocol, one must use a symmetric system in which
the two lasers have symmetric dynamics \cite{Noam}. We therefore
wish to discuss the symmetry of the synchronized states. The
symmetry of the isochronal synchronization is nearly perfect, but
the symmetry of the achronal synchronization, with $\kappa=0$,
needs further investigation, as the correlation displayed in Fig.
\ref{fb_hist} is averaged. In our numerical simulations we observe
that between the LFF breakdowns there is a high cross correlation
both with delay $+\tau_{c}$ and $-\tau_{c}$, even without
averaging over segments, as if the leading role is shared by the
the two lasers. Therefore in the areas between the breakdowns
there is considerable symmetry in the laser intensity sequences,
whereas at the breakdowns, this symmetry is broken and the lasers
exchange the leading role randomly between them i.e. sometimes $A$
falls before $B$ and sometimes vice versa. Fig.
\ref{histo_achronal} displays a histogram of the ratio of the
cross correlation with delay $+\tau_{c}$ and with delay
$-\tau_{c}$, denoted as $\frac{\rho(+\tau_{c})}{\rho(-\tau_{c})}$.
The peak at $1$ indicates a high symmetry throughout the sequence.
The inset of Fig. \ref{histo_achronal} displays a histogram of the
time delay between the breakdowns of laser $A$ and $B$. The two
peaks at $\pm \tau_{c}$ indicate that about half of the falls are
lead by laser $A$ and the other half by laser $B$. We conclude
from these results that the achronal synchronization might be
suitable for cryptographic purposes when excluding the LFF
breakdowns, or in parameter-regimes in which LFF breakdowns do not
appear but the signal is still chaotic.

It is evident from the discussion above that for two mutually
coupled lasers without self feedback ($\kappa=0$), isochronal
synchronization is not stable, yet by merely adding self feedback,
isochronal synchronization becomes stable. This appears for a wide
window of parameters. We do not prove here the stability or
measure the distribution of Lyapunov exponents, but rather give an
intuitive explanation for the difference between a system with and
without self feedback.

In a system without self feedback each laser receives the delayed
signal of the other laser. Starting from different initial
conditions, even if after some time the lasers come very close to
each other, they still have a different history which prevents
them from completely synchronizing, because each laser continues
to receive a different signal, $E_{A}(t-\tau_{c}) \ne
E_{B}(t-\tau_{c})$. Only if the two lasers reach a state in which
their optical field and phase over an entire window of size
$\tau_{c}$ is close enough, will they manage to remain
synchronized. Otherwise, their different histories will drive them
to achronal synchronization.

In a system with $\kappa>0$ on the other hand, each laser receives
the delayed signals of $both$ lasers. Even if their history is
different, the two lasers receive the same signal. It is then
easier for them to remain synchronized because the difference in
their delayed values does not affect the synchronization.

Another way to put this argument is by looking at one of the
necessary conditions for isochronal synchronization. In order for
isochronal synchronization to exist, the condition that
$\frac{dE_{A}(t)}{dt}=\frac{dE_{B}(t)}{dt}$ must be satisfied,
because the lasers' dynamics are first order differential
equations. When $\kappa=0$ and the lasers start in different
initial states, this condition is only satisfied if $E_{A}(t) =
E_{B}(t)$ holds for every time $t$ over the interval
$[t,t+\tau_{c}]$. However, when $\kappa>0$ and
$\kappa_{A}+\sigma_{A}=\kappa_{B}+\sigma_{B}$, then
$\frac{dE_{A}(t)}{dt}=\frac{dE_{B}(t)}{dt}$ immediately follows.
Therefore, a necessary condition for synchronization is more
easily fulfilled for $\kappa>0$.


In conclusion, the existence of a stable symmetric isochronal
synchronization has been demonstrated in coupled lasers with
self-feedback.  In coupled lasers without self-feedback we have
shown that partial symmetry appears in the form of achronal
synchronization. The synchronization methods open the possibility
of implementing optically chaotic systems in different
communication tasks.


\begin{thebibliography}{9}
\bibitem{remark2}The absolute values of $\sigma$ and $\kappa$ were difficult to
attain experimentally but we were able to give a good estimation
as to the rate of self feedback ($\kappa$) relative to the
coupling rate ($\sigma$) by comparing the effect they have on the
solitary laser. When subjected to each of them alone, it shows an
increase in its output power.
\bibitem{remark1}Note that the intermediate areas in which the isochronal synchronization is
less stable, dissapear in the case of $\kappa_{A}=\sigma_{B}$ and
$\kappa_{B}=\sigma_{A}$ and instead there is stable isochronal
synchronization for every $\kappa>0$.
\end{thebibliography}
\end{document}